\documentclass[preprint,
aps,showpacs,tightenlines,
nofootinbib,showkeys,byrevtex]{revtex4}

\usepackage{graphicx}%
\usepackage{dcolumn}
\usepackage{amsmath}
\usepackage{epsfig}
\usepackage{latexsym}

\def\darr#1{\raise1.5ex\hbox{$\leftrightarrow$}\mkern-16.5mu #1}
\def\){\right)} 
\def\({\left(} 
\def\]{\right]} 
\def\[{\left[}

\def\Align{&=&}

\def\CPT{$\chi${\rm PT}}
\def\dslash{{D\hskip-0.65em /}}
\def\CSW{c_{\rm SW}}
\def\si{{}^1\kern-.14em S_0}
\def\siii{{}^3\kern-.14em S_1}
\def\diii{{}^3\kern-.14em D_1}

\def\pone{{}^3\kern-.14em P_1}
\def\pzero{{}^3\kern-.14em P_0}
\def\ptwo{{}^3\kern-.14em P_2}

\def\nrcpt{NR\raise.4ex\hbox{$\chi$}PT\ }

\def\ltap{\ \raise.3ex\hbox{$<$\kern-.75em\lower1ex\hbox{$\sim$}}\ }
\def\gtap{\ \raise.3ex\hbox{$>$\kern-.75em\lower1ex\hbox{$\sim$}}\ }

\newcommand{\eqn}[1]{\label{eq:#1}}
\newcommand{\refeq}[1]{(\ref{eq:#1})} 
\newcommand{\eq}{eq.~\refeq}

\newcommand{\Eqsii}[2]{Eqs.~(\ref{eq:#1}),~(\ref{eq:#2})}
\newcommand{\Eq}{Eq.~\refeq} 
\newcommand{\Eqs}[2]{Eqs.~\refeq{#1}-\refeq{#2}}
\newcommand{\beq}{\begin{eqnarray}}% can be used as {equation} or {eqnarray}
\newcommand{\eeq}{\end{eqnarray}}

\newcommand{\mcal}[1]{{\mathcal #1}}
\def\Journal#1#2#3#4{{#1} {\bf #2}, #3 (#4)}

\def\NPB{{\em Nucl. Phys.} B}

\def\PRD{{\em Phys. Rev.} D}

\begin{document}
\bibliographystyle{unsrt}     
\preprint{TRI-PP-01-38}
\preprint{BUHEP-02-04}

\title{Chiral perturbation theory for the Wilson lattice action}

\author{Gautam Rupak~${}^{a,\,b,\,}$\footnote{Email: {\tt grupak@triumf.ca}; 
permanent address: $(b)$} 
and
Noam Shoresh~${}^{c,\,d,\,}$\footnote{Email: {\tt shoresh@bu.edu};
permanent address: $(d)$}
}
\affiliation{${}^a$TRIUMF, Vancouver, B.C., Canada V6T 2A3\\
${}^b$Lawrence Berkeley National Laboratory, Berkeley, CA, U.S.A. 94720 \\
${}^c$Department of Physics, University of Washington, 
Seattle, WA, U.S.A. 98195 \\ 
${}^d$Department of Physics, Boston University, Boston, WA, U.S.A. 02215}

%\date{\today}% It is always \today, today, but you may specify any date with \date.

\begin{abstract}
We extend chiral perturbation theory to include linear 
 dependence on the lattice spacing $a$ for the Wilson action. The perturbation
theory is written as a double expansion in the 
small quark mass $m_q$ and lattice 
spacing $a$. We present formulae for the 
mass and decay constant of a flavor-non-singlet meson in this scheme to order $a$ and $m_q^2$. 
The extension to the partially quenched theory is also described.

\end{abstract}

% PACS, the Physics and Astronomy Classification Scheme.
%Use showkeys class option if keyword
                              %display desired
\maketitle 
%\nopagebreak[4]

%%%%%%%%%%%%%%%%%%%%%%%%%%%%%%%%%%%%%%%%%%%%%%%%%%%%%%%%%%%%%
\begin{section}{Introduction}
\label{introduction}
%%%%%%%%%%%%%%%%%%%%%%%%%%%%%%%%%%%%%%%%%%%%%%%%%%%%%%%%%%%%
Chiral perturbation theory (\CPT) is an important tool for extracting
quantitative information from lattice 
simulations of QCD. The reason for this is that it is impractical to have
dynamical quarks in simulations that are as light as the up and down quarks,
and \CPT~is needed for a controlled, systematic 
extrapolation in the quark masses. 
Since \CPT~describes {\em continuum} QCD at low-energies, its
application in numerical simulations is possible only after
extrapolating lattice data to the continuum limit where the lattice spacing,
$a$, vanishes.
In this paper, we study the behavior of the Wilson lattice action
close to the continuum
by incorporating $\mcal O(a)$ effects in a reformulation
of \CPT. A similar approach was first taken in 
Ref.~\cite{SharpeSingleton1998} to
investigate the phase diagram for Wilson fermions in two-flavor QCD.

The quark mass matrix (considering only the 2 or 3 lightest quarks) has a
special role in QCD --- it parameterizes the explicit breaking of the axial
symmetries. As a result, the light quark masses appear explicitly in the 
low-energy effective theory. In this paper, we exploit the fact that for the
Wilson lattice action there is another independent symmetry breaking
parameter, linear in
$a$~\cite{SharpeSingleton1998,Luscher}. To $\mcal{O}(a)$ this is the
only discretization effect, and thus a generalization of the chiral
Lagrangian can be written which includes all terms linear in $a$.

%%%%%%%%%%%%%%%%%%%%%%%%%%%%%%%%%%%%%%%%%%%%%%%%%%%%%%%%%%%%%
\end{section}
\begin{section}{Effective Lagrangian}
\label{EFTSection}
%%%%%%%%%%%%%%%%%%%%%%%%%%%%%%%%%%%%%%%%%%%%%%%%%%%%%%%%%%%%%

The Wilson action for fermions is given by
\beq\eqn{latAct}
\mcal S_{F}^{(W)}\Align 
\sum_x\left[\bar\psi(x)\gamma_\mu\Delta_\mu(x)\psi(x)
+\bar\psi(x) m_q\psi(x)
+a r \bar\psi(x) \Delta^2\psi(x)\right]\ ,
\eeq
where 
\beq
\Delta_\mu\psi(x)=\frac{1}{2a}\left[U_\mu(x)\psi(x+a\hat\mu)
        -U_\mu^\dagger(x-a\hat\mu)\psi(x-a\hat\mu)\right].
\eeq
$U_\mu(x)$ is the gauge group valued field
defined on the links of the lattice. 
In typical lattice simulations, $r=1$ but we will keep it more general
for now. 
Since we are interested in a perturbative study of discretization
effects, we follow \cite{Symanzik83} and consider an
effective action in the continuum which describes the same physics as the
discrete lattice action (including the gauge action), 
well below the cutoff $1/a$. The effective action
is expanded in powers of $a$:
\beq\eqn{continuum}
S_{\mbox{eff}}\Align S_0+a S_1 + a^2 S_2+\cdots\ .
\eeq
By construction, $S_0$ is the QCD action. 
Symmetry considerations restrict the number of mass-dimension $5$ operators
that appear in $S_1$. The equations of motion can be used to further reduce
the list of operators. One then identifies the operators that already appear
in $S_0$, which give rise to the renormalization of the quark masses and the
coupling constant. Finally, one is left with a single new term at
$\mcal O(a)$, the 
Pauli term~\cite{Luscher}:
\beq
aS_1\Align a\CSW \bar\psi\sigma_{\mu\ \nu} F^{\mu\ \nu}\psi\ .
\eqn{Pauli}
\eeq
$\CSW$ is a constant of $\mcal{O}(1)$ which is a complicated function of the
gauge coupling and $r$.

Considering $S_0+aS_1$ as an underlying theory, it has an $SU(N_f)_L\times
SU(N_f)_R$ flavor symmetry which is broken down to the vector part by the
mass and Pauli terms with coefficients $m_q$ and $a\CSW$ respectively. $\bar\psi\psi$ and
$\bar\psi\sigma_{\mu\ \nu} F^{\mu\ \nu}\psi$ break the axial symmetry in the
same way and therefore, from a spurion analysis point of view, $m_q$ and
$a\CSW$ are on equal footing, as noted in Ref.~\cite{SharpeSingleton1998}.
 In particular, $a\CSW$ is treated as a
non-trivial matrix in flavor space, just like $m_q$. It is possible to
give $a\CSW$ a flavor structure in simulations by promoting the constant $r$
to a matrix. This can supply extra ``knobs'' which can aid the
continuum extrapolations. The squares of
the Goldstone boson (GB) masses are linear in the symmetry breaking
matrices of parameters~\cite{SharpeSingleton1998},
 conveniently written as:
\beq
\eqn{chidef}
\chi\equiv 2 B_0 m_q\ , 
\hspace{0.5in}\rho\equiv 2 W_0 a\CSW\ .
\eeq 
$B_0$ and
$W_0$ are unknown dimensionful parameters that appear in the effective
Lagrangian at leading order (LO) which is defined below.

\CPT~for the Wilson
action (which will be denoted by W\CPT) is an expansion in the squares of
the small momenta $p$ and
the small pseudo Goldstone boson masses.    
We 
formally consider the expansion to be in
two independent small parameters:
\beq
\epsilon\sim\frac{p^2}{\Lambda_\chi^2}\sim\frac{\chi}{\Lambda_\chi^2}\quad\mbox{
  and }\quad\delta\sim\frac{\rho}{\Lambda_\chi^2}.
\eeq
$\Lambda_\chi$ is the scale where new high-energy 
physics enters and the effective field
theory (EFT) 
no longer describes the correct physics. This happens around the mass
of the rho meson, or roughly $\Lambda_\chi\sim 1\mathrm{GeV}$.
From dimensional analysis, and the fact that $B_0$ and $W_0$ depend only on
the high-energy details, one can check that the expansion is in fact in
$m_q/\Lambda_\chi$ and in $a\Lambda_\chi$.

Since \Eq{continuum} 
is truncated at $\mcal{O}(a)$ it makes no sense to go
beyond $\mcal{O}(\delta)$ in W\CPT~(we remark on $\mcal{O}(a^2)$ corrections in
the next 
section). For convenience, we choose to collect terms in the
LO and next-to-leading order (NLO) Lagrangians as follows:
\begin{align}
 \text{LO}: \hspace{.5in}& 
 \mcal{L}_2 \sim \mcal{O}\left( {\epsilon ,\delta } \right)\ , \\ 
  \text{NLO}:  \hspace{.5in} &
\mcal{L}_4 \sim\mcal{O}\left( {\epsilon ^2 ,\epsilon \delta } \right)\ .
\end{align}
The underlying hierarchy consistent with this choice is
%\begin{align}
$\left\{ {\epsilon ,\delta } \right\} \gg \left\{ {\epsilon ^2 ,\epsilon \delta } \right\} \gg \delta ^2$, 
%\end{align}
and the last inequality also implies $\epsilon\gg\delta$. In practice,
the perturbative expansion should be organized according to the actual
sizes of the expansion parameters, which are determined by the quark
masses in the simulation and the size of the lattice
spacing.
For example,
if the simulations are done so close to the continuum that $\delta$ is very
small, it might make more sense to have the LO Lagrangian be
$\mcal{O}(\epsilon)$, and at NLO $\mcal{O}(\epsilon^2,\delta)$.
With our convention,  
\beq\eqn{L2}
\mathcal{L}_2  \Align \frac{{f^2 }}
{4}\operatorname{tr} (\partial \Sigma\partial \Sigma^\dag  ) - \frac{{f^2 }}
{4}\operatorname{tr} ((\chi  + \rho )\Sigma^\dag   
+ \Sigma(\chi ^\dag   + \rho ^\dag 
 ))\ ,
\eeq
is the LO Lagrangian, where $\Sigma=\exp(2 i\Pi/f)$
 contains the matrix of meson fields, $\Pi$.
It is useful to note that \Eq{L2} can be ``produced'' from the LO
Lagrangian of ordinary \CPT~by the substitution $\chi\to\chi+\rho$.

The NLO
Lagrangian is:
\beq\eqn{L4} 
\mcal L_4\Align
{L_1 \left\langle {\partial \Sigma\partial \Sigma^\dag  } \right\rangle ^2 }+
{L_2 \left\langle {\partial _\mu  \Sigma\partial _\nu  \Sigma^\dag  }
  \right\rangle 
 \left\langle {\partial _\mu  \Sigma\partial _\nu  \Sigma^\dag  }
 \right\rangle }+ 
{L_3 \left\langle {(\partial \Sigma\partial \Sigma^\dag  )^2 }
  \right\rangle }\nonumber\\ 
&{}&+{L_4 \left\langle {\partial \Sigma\partial \Sigma^\dag  }
 \right\rangle \left\langle 
{\chi^\dag  \Sigma + \Sigma^\dag  \chi } \right\rangle }
+{W_4 \left\langle {\partial \Sigma\partial \Sigma^\dag  }
  \right\rangle \left\langle  
{\rho ^\dag  \Sigma + \Sigma^\dag  \rho } \right\rangle }\nonumber\\
&{}&+{L_5 \left\langle {\partial \Sigma\partial \Sigma^\dag 
 (\chi ^\dag  \Sigma + \Sigma^\dag  \chi )} \right\rangle }
+{W_5 \left\langle {\partial \Sigma\partial \Sigma^\dag  (\rho ^\dag  
\Sigma + \Sigma^\dag  \rho )} \right\rangle }\nonumber\\
&{}&+{L_6 \left\langle {\chi ^\dag  \Sigma + \Sigma^\dag  \chi }
  \right\rangle ^2 } 
+{W_{6} \left\langle {\chi ^\dag  \Sigma + \Sigma^\dag  \chi }
  \right\rangle \left\langle {\rho ^\dag  \Sigma + \Sigma^\dag  \rho }
  \right\rangle }\nonumber\\ 
&{}&+ {L_7 \left\langle {\chi ^\dag  \Sigma - \Sigma^\dag  \chi }
  \right\rangle ^2 } 
+{W_{7} \left\langle {\chi ^\dag  \Sigma - \Sigma^\dag  \chi } \right\rangle 
\left\langle {\rho ^\dag  \Sigma - \Sigma^\dag  \rho } \right\rangle
}\nonumber\\ 
&{}&+{L_8 \left\langle {\chi ^\dag  \Sigma\chi ^\dag  \Sigma + \Sigma^\dag  
\chi \Sigma^\dag  \chi } \right\rangle }
+{W_{8} \left\langle {\rho ^\dag  \Sigma\chi ^\dag  \Sigma + \Sigma^\dag 
 \rho \Sigma^\dag  \chi } \right\rangle }\ .
\eeq
The angled brackets stand for traces over the flavor indices. In the
limit $a\rightarrow 0$, the  
$L_i$'s above are the usual Gasser-Leutwyler (GL) coefficients of \CPT.

A word  about  certain  $\log(a\Lambda)$ corrections  is  appropriate
here. In  the EFT formulation, the  Lagrangian is written  in terms of
the most general set of  operators constructed out of the relevant degrees
of freedom that respect the symmetries of the theory.  The high-energy
physics that  was integrated out enters through  the unknown couplings
that  multiply  these operators.   Thus  the  low-energy constants  or
couplings are entirely determined  by the high-energy scales. In \CPT,
the operators contain only the  light meson and photon fields, and the
low-energy constants $B_0$, $f$, $L_i$, etc., are functions of the QCD
scale $\Lambda_{\rm QCD}$. In  particular, the $L_i$'s are independent
of the pion masses  $m_\pi\sim\sqrt{m_q}$ which are associated with the
long-distance physics. All the  $m_q$ dependence of \CPT~is explicitly
written in the  operators.  This still holds for  the $m_q$ dependence
of the W\CPT~Lagrangian written above, but the same cannot be said about
the lattice spacing  $a$. It is true that an  $\mcal O(a)$ term breaks
the chiral  symmetry in the same  way as $m_q$, and  $a\Lambda^2$ is a
soft scale  associated with  the pseudo  GB mass, but  $1/a$ is  not a
soft scale --- it acts as the ultra-violet cutoff for the  discrete lattice.
Thus,  while  the 
low-energy constants of  W\CPT~ are  expected to be  independent of $m_q$ 
 and $a\CSW$, they  could in principal  have a complicated dependence  on the 
gauge coupling $g$ (in $S_0$), which itself depends on $a$.
However, the  running of  the  coupling
constant is  determined in simulations 
by requiring that  as one approaches the continuum, some
chosen physical  quantity remains fixed. Effectively,  this means that
the coupling $g$ and 
the cutoff $1/a$ combine to give the only real scale in the
theory  -- $\Lambda_{QCD}$ --  and the  continuum limit  is approached
smoothly.   We therefore  expect  the $L_i$'s  and
$W_i$'s to depend on $\Lambda_{QCD}$, and only weakly on $a$, the latter 
dependence coming from
higher orders  in perturbation theory,  or involving higher  powers of
$a$  which can  always be  expanded.\footnote{We thank Paulo Bedaque
and  Andrew Cohen  for helping us understand this issue.}
The parameter  $\CSW$ in  the
action  $S_1$ will  still depend  on $\log  (a\Lambda_{QCD})$,  and it
might  be possible  to calculate  these  dependences  explicitly in
perturbation close  to the continuum~\cite{LuscherWeisz96,BGLS2001}.

A related form of implicit $a$ dependence exists in the quark masses. The quark
masses that appear in the W\CPT~Lagrangian (\Eqsii{L2}{L4})  
are not the same
as those that appear in the Wilson action (\Eq{latAct}). Because of the
explicit breaking of chiral symmetry due to the Wilson term, the quark masses
are not protected from additive renormalization of the order of the lattice
cutoff $1/a$. In practice one finds in simulations a ``critical'' line
$m_q^c(a)$ on which the meson masses approximately vanish. The quark mass is
then defined as the distance from this line:
\beq
\tilde m_q=m_q-m_q^c(a),
\eeq
and it is $\tilde m_q$ that should be used in \Eqsii{L2}{L4}.
$\tilde m_q$ compensates for the large
$\mcal{O}(1/a)$ shift in the quark masses, but it also contains
positive powers of $a$. 
This is not a problem --- redefinitions of the mass parameter of
this sort only lead to changes in the $W_i$'s. The GL
coefficients are not affected because the operators with which they are
associated do not contain $a$. The re-shuffling of the $W_i$'s does mean,
however, that their actual numerical values depend on the prescription that
is used to determine $m_q^c(a)$ and to define $\tilde m_q$.

%\footnote{
Note 
  that the chiral limit cannot be taken by simply setting $\tilde m_q\to
  0$. While $m_q^c$ satisfies $M_\pi^2(m_q^c(a),a)=0$, there is no reason
  that other quantities will attain their chiral limit for this value of
  $m_q$. This is a reflection of the fact that there really are 2 different 
  operators that break the symmetry.
%}

%%%%%%%%%%%%%%%%%%%%%%%%%%%%%%%%%%%%%%%%%%%%%%%%%%%%%%%%%%%%%
\end{section}
\begin{section}{Applications}
\label{applications}
%%%%%%%%%%%%%%%%%%%%%%%%%%%%%%%%%%%%%%%%%%%%%%%%%%%%%%%%%%%%%

In the following two
subsections we
calculate the expressions for the mass and decay constant of a
flavor-charged meson with the
flavor indices $AB$ ($A\ne B$) having the same quantum numbers as
$\bar\psi_B\gamma_5\psi_A$. 
 In
the calculations that follow we take $\chi$ to be a
diagonal matrix with entries $(\chi)_{ii}=\chi_i$, and use the notation
$\chi_{AB}=(\chi_{A}+\chi_{B})/2$. Note that this
notation coincides with the standard way of denoting matrix elements
only for the diagonal entries. The same convention is used for
$\rho$. It is convenient to define another matrix, $\mu=\chi+\rho$,
which is the combination that appears in $\mcal{L}_2$. The
subscript notation for $\mu$ follows that of $\chi$ and $\rho$, except
for the quantities $\mu_\pi$ and $\mu_\eta$ which are defined below.

%%%%%%%%%%%%%%%%%%%%%%%%%%%%
\begin{subsection}{Masses}
\label{mass}
%%%%%%%%%%%%%%%%%%%%%%%%%%%%
The mass of a flavor-charged meson in W\CPT~with three
quark flavors is given
through NLO by
\beq
M_{AB}^2=(M_{AB}^2)_\mathrm{LO}+(M_{AB}^2)_\mathrm{NLO,\
  loop}+(M_{AB}^2)_\mathrm{NLO,\ tree}\ ,
\eeq
with
\begin{align}
\left( {M_{AB}^2 } \right)_{\text{LO}}  =  & \mu _{AB} \eqn{WmassLO}\ , \\ 
  \left( {M_{AB}^2 } \right)_{\text{NLO, loop}}  =  & \frac{1}
{{48f^2 \pi ^2 }}\,\mu _{AB} \sum\limits_{x = \pi ,\eta } {R_x^{AB} \mu _x \log
  \mu _x } \eqn{WmassNLOloop}\ , \\ 
  \left( {M_{AB}^2 } \right)_{\text{NLO, tree}}  =  &  - \frac{{24}}
{{f^2 }}L_4 \left( {\chi _{AB}  + \rho _{AB} } \right)\bar \chi  - \frac{{24}}
{{f^2 }}W_4 \chi _{AB} \bar \rho  - \frac{8}
{{f^2 }}L_5 \left( {\chi _{AB}  + \rho _{AB} } \right)\chi _{AB} \notag  \\ 
   &  - \frac{8}
{{f^2 }}W_5 \chi _{AB} \rho _{AB}  + \frac{{24}}
{{f^2 }}2L_6 \chi _{AB} \bar \chi  + \frac{{24}}
{{f^2 }}W_6 \left( {\chi _{AB} \bar \rho  + \rho _{AB} \bar \chi } \right)\notag  \\ 
   &  + \frac{{24}}
{{f^2 }}2L_8 \chi _{AB}^2  + \frac{8}
{{f^2 }}2W_8 \chi _{AB} \rho _{AB} \,,\eqn{WmassNLOtree}
\end{align}
% MathType!End!2!1!
where $\mu_\pi$
and $\mu_\eta$ are the squares of the LO masses of
the two light flavor-neutral mesons, given implicitly by
% MathType!Translator!2!1!Noam LaTeX.tdl!Noam -- AMS-LaTeX!
% MathType!MTEF!2!1!+-
% feaafaart1ev1aaatCvAUfeBSjuyZL2yd9gzLbvyNv2CaerbuLwBLn
% hiov2DGi1BTfMBaeXatLxBI9gBaerbd9wDYLwzYbItLDharqqtubsr
% 4rNCHbGeaGqiVu0Je9sqqrpepC0xbbL8F4rqqrFfpeea0xe9Lq-Jc9
% vqaqpepm0xbba9pwe9Q8fs0-yqaqpepae9pg0FirpepeKkFr0xfr-x
% fr-xb9adbaqaaeGaciGaaiaabeqaamaabaabaaGceaqabeaacqaH8o
% qBdaWgaaWcbaGaeqiWdahabeaakiabgUcaRiabeY7aTnaaBaaaleaa
% cqaH3oaAaeqaaOGaeyypa0JaaGOmaiqbeY7aTzaaraaabaGaeqiVd0
% 2aaSbaaSqaaiabec8aWbqabaGccqaH8oqBdaWgaaWcbaGaeq4TdGga
% beaakiabg2da9iabeY7aTnaaBaaaleaacaaIXaaabeaakiabeY7aTn
% aaBaaaleaacaaIYaaabeaakiabgUcaRiabeY7aTnaaBaaaleaacaaI
% XaaabeaakiabeY7aTnaaBaaaleaacaaIZaaabeaakiabgUcaRiabeY
% 7aTnaaBaaaleaacaaIYaaabeaakiabeY7aTnaaBaaaleaacaaIZaaa
% beaakiaaykW7caGGUaWaaObaaeaacaWGxbGaamiCaiaadMgacaWGLb
% GaamiDaiaadggaaeqaaaaaaa!6385!
\begin{align}
& \mu _\pi   + \mu _\eta   = 2\bar \mu\ ,  \\ 
 & \mu _\pi  \mu _\eta   = \mu _1 \mu _2  + \mu _1 \mu _3  + \mu _2 \mu _3 \,.\eqn{Wpieta}
\end{align}
% MathType!End!2!1!
Here $\bar{\chi}=tr(\chi)/3$ and similarly for $\rho$ and $\mu$. 
Also, if we denote by $C$ the flavor that is different from both
$A$ and $B$, we have
% MathType!Translator!2!1!Noam LaTeX.tdl!Noam -- AMS-LaTeX!
% MathType!MTEF!2!1!+-
% feaafaart1ev1aaatCvAUfeBSjuyZL2yd9gzLbvyNv2CaerbuLwBLn
% hiov2DGi1BTfMBaeXatLxBI9gBaerbd9wDYLwzYbItLDharqqtubsr
% 4rNCHbGeaGqiVu0Je9sqqrpepC0xbbL8F4rqqrFfpeea0xe9Lq-Jc9
% vqaqpepm0xbba9pwe9Q8fs0-yqaqpepae9pg0FirpepeKkFr0xfr-x
% fr-xb9adbaqaaeGaciGaaiaabeqaamaabaabaaGcbaGaamOuamaaDa
% aaleaacqaHapaCaeaacaWGbbGaamOqaaaakiabg2da9maalaaabaGa
% eqiVd02aaSbaaSqaaiaadoeaaeqaaOGaeyOeI0IaeqiVd02aaSbaaS
% qaaiabec8aWbqabaaakeaacqaH8oqBdaWgaaWcbaGaeq4TdGgabeaa
% kiabgkHiTiabeY7aTnaaBaaaleaacqaHapaCaeqaaaaakiaacYcaca
% aMf8UaamOuamaaDaaaleaacqaH3oaAaeaacaWGbbGaamOqaaaakiab
% g2da9maalaaabaGaeqiVd02aaSbaaSqaaiaadoeaaeqaaOGaeyOeI0
% IaeqiVd02aaSbaaSqaaiabeE7aObqabaaakeaacqaH8oqBdaWgaaWc
% baGaeqiWdahabeaakiabgkHiTiabeY7aTnaaBaaaleaacqaH3oaAae
% qaaaaakmaaAaaabaGaam4vaiaadkfaaeqaaiaaykW7caGGUaaaaa!65E3!
\begin{align}
R_\pi ^{AB}  = \frac{{\mu _C  - \mu _\pi  }}
{{\mu _\eta   - \mu _\pi  }},\quad R_\eta ^{AB}  = \frac{{\mu _C  - \mu _\eta  }}
{{\mu _\pi   - \mu _\eta  }}\eqn{WR}\,.
\end{align}
% MathType!End!2!1!

In deriving the expressions for the mass in W\CPT~one could use a very
convenient ``trick'' relating these expressions to the corresponding
expressions 
in ordinary \CPT. 
As mentioned earlier, the LO Wilson chiral 
Lagrangian $\mcal L_2$, \Eq{L2}, can be obtained from \CPT~LO 
Lagrangian by the simple  
substitution $\chi\rightarrow\chi +\rho$ (or $\chi\to\mu$). Thus any quantity
$h(\chi)$ in \CPT~that depends only on the LO Lagrangian can be
trivially reproduced in W\CPT~according to $h(\chi)\rightarrow h(\chi+\rho)$. 
This is true for the LO and NLO loop diagrams that contribute to the
mass. Similar results also hold for the decay constant. 
We provide the expressions for the mass in ordinary \CPT~in
Appendix~\ref{ordinaryCPT} for comparison.

%%%%%%%%%%%%%%%%%%%%%%%%%%%%%%%%%%%%%%%%%%%%%%%%%%%%%%%%%%%%
\end{subsection}
\begin{subsection}{Decay constants}
\label{decay}
%%%%%%%%%%%%%%%%%%%%%%%%%%%%%%%%%%%%%%%%%%%%%%%%%%%%%%%%%%%%

The decay constant is given through NLO by:
\beq
f_{AB}
\Align
(f_{AB})_{\text{LO}}+(f_{AB})_{\text{NLO, loop}}
+(f_{AB})_{\text{NLO, tree}}\ , 
\eeq
with
\begin{align}
\left( {f_{AB} } \right)_{\text{LO}}  =  & f\eqn{WfLO}\ , \\ 
  \left( {f_{AB} } \right)_{\text{NLO, loop}}  =  &  - \frac{1}
{{64\pi ^2 f}}\sum\limits_{\begin{subarray}{l} 
  i = 1,2,3 \\ 
  j = A,B 
\end{subarray}}  {\mu _{ij} \log \mu _{ij} }  + \frac{1}
{{192\pi ^2 f}}\left( {\mu _A  - \mu _B } \right)\left\{ {\log \left( {\frac{{\mu _A }}
{{\mu _B }}} \right) + } \right.\notag \eqn{WfNLOloop} \\ 
   & \left. { + \sum\limits_{x = \pi ,\eta } {R_x^{AB} \mu _x \left[ {\frac{{\log \left( {{{\mu _A } \mathord{\left/
 {\vphantom {{\mu _A } {\mu _x }}} \right.
 \kern-\nulldelimiterspace} {\mu _x }}} \right)}}
{{\mu _A  - \mu _x }} - \frac{{\log \left( {{{\mu _B } \mathord{\left/
 {\vphantom {{\mu _B } {\mu _x }}} \right.
 \kern-\nulldelimiterspace} {\mu _x }}} \right)}}
{{\mu _B  - \mu _x }}} \right]} } \right\}\ , \\ 
  \left( {f_{AB} } \right)_{\text{NLO, tree}}  =  & \frac{{12}}
{f}\left( {L_4 \bar \chi  + W_4 \bar \rho } \right) + \frac{4}
{f}\left( {L_5 \chi _{AB}  + W_5 \rho _{AB} } \right)\,.\eqn{WfNLOtree}
\end{align}
% MathType!End!2!1!

%%%%%%%%%%%%%%%%%%%%%%%%%%%%%%%%%%%%%%%%%%%%%%%%%%%%%%%%%%%%
\end{subsection}
\begin{subsection}{W\CPT,  $\mcal{O}(a^2)$, and improvement}
%%%%%%%%%%%%%%%%%%%%%%%%%%%%%%%%%%%%%%%%%%%%%%%%%%%%%%%%%%%%

In the simplest sense, the expressions for the mass and decay
constant in W\CPT~can be used to aid
in taking the continuum limit. These forms provide all the linear $a$
dependence, as well as non-trivial logarithms that involve $a$ and
$m_q$. A test of these formulae would be to check whether they describe the
$a$ dependence better than naive extrapolations. Perhaps a more useful way to
think about it is that with these 
expressions one can determine the GL coefficients directly from lattice data
at finite $a$. 

What about higher orders in $a$? At order $a^2$ the picture changes
qualitatively. There are operators in $S_2$, such as $\bar\psi\dslash D_\mu
D_\mu\psi$, that do not break the chiral symmetry. 
This means that $a$ can no
longer be associated only with symmetry breaking effects, and spurion
analysis cannot be used to constrain the $a^2$ operators. Nevertheless, we
might still expand in $\epsilon$ and $\delta$ simultaneously. The
LO, $\mcal{O}(\epsilon,\delta)$ Lagrangian, and consequently 
the LO mass and decay
constant are unchanged. At NLO,
$\mcal{O}(\epsilon^2,\epsilon\delta,\delta^2)$, there are several 
$\mcal O(a^2)$ 
operators that are added to the Lagrangian, but they are all independent of
the quark masses and do not contain derivatives. Consequently, the only correction to the meson masses at this
order is an additional term of the form $\omega a^2$, where $\omega$ is an
unknown constant of mass-dimension 4. The expression for the decay constant
does not receive {\em any} corrections at this order. This is because
tree level 
contributions to the decay constant can only come from operators that contain
derivatives. 

Improvement schemes (first suggested by Symanzik in Ref.\cite{Symanzik83}) 
--- using an improved action and improved operators --- are
another important tool for studying and reducing discretization effects.
Using improved action involves adding a discretized version of
the Pauli term in \Eq{Pauli} to the Wilson action in 
\Eq{latAct} which exactly cancels the $S_1$ term in the continuum action, 
\Eq{continuum}.
This means that up to $\mcal{O}(a^2)$ the lattice theory is just QCD,
and at low-energies we expect \CPT~to be a good description. From the
perspective of W\CPT~this is equivalent to saying that using an
improved action sets all $W_i=0$. This is of course not surprising:
the use of improved action is meant to eliminate all $\mcal{O}(a)$
dependence from observables, and the $W_i$ coefficients parameterize
exactly this dependence. One should not conclude from this
  that improving the action is enough for complete $\mcal{O}(a)$
  improvement. As mentioned above, some dimension-5 operators 
that are allowed by the symmetries, and can therefore appear in $S_1$,
are implicitly absorbed in $S_0$ by replacing the bare parameters of
$S_0$ with renormalized ones.
This is a necessary step, and it is compatible with the fact that in
improvement schemes, in addition to using an improved action, one must 
use improved operators.

%%%%%%%%%%%%%%%%%%%%%%%%%%%%%%%%%%%%%%%%%%
\end{subsection}
\end{section}
\begin{section}{Partially quenched theories}
\label{Pquenched}
%%%%%%%%%%%%%%%%%%%%%%%%%%%%%%%%%%%%%%%%%%

W\CPT~ is appropriate for the type of lattice simulations which are
called ``unquenched''.
These are simulations in which there are 2 or 3 dynamical
fermions (also called ``sea quarks''), and 
expectation values are calculated of operators which are constructed
from a different type of fermions (``valence quarks'') which have
the same masses as the sea quarks. 
In most lattice simulations, however,
the masses of the valence quarks are not taken to be the 
same as those of the sea quarks. Simulations that are done this way
are called partially 
quenched (PQ).
Theoretically this can
be described 
by a QCD-like construction which includes
ghosts~\cite{morel87,Bernard}. The low-energy behavior of these
theories is described by PQ 
\CPT~\cite{Bernard}, which has the same unknown low-energy constants
as \CPT~ for ordinary QCD~\cite{SharpeShoresh2001}. 
Thus, PQ simulations provide additional mass parameters that
can be used to probe the theory in a larger parameter space, and gain
better statistics in determining the GL coefficients
\cite{SharpeShoresh2000,CohenKaplanNelson99}. It 
is of clear practical value to consider the generalization of W\CPT~
to the PQ case.

PQ QCD contains three different types of spin-half particles --- valence
quarks, sea quarks, and ghosts which obey Bose-Einstein
statistics. There is a single ghost flavor for every valence quark, and
they both have the same mass. The quark mass matrix for a theory with
2 valence quarks, 3 sea quarks, and 2 ghosts is
% MathType!Translator!2!1!Noam LaTeX.tdl!Noam -- AMS-LaTeX!
% MathType!MTEF!2!1!+-
% feaafaart1ev1aaatCvAUfeBSjuyZL2yd9gzLbvyNv2CaerbuLwBLn
% hiov2DGi1BTfMBaeXatLxBI9gBaerbd9wDYLwzYbItLDharqqtubsr
% 4rNCHbGeaGqiVu0Je9sqqrpepC0xbbL8F4rqqrFfpeea0xe9Lq-Jc9
% vqaqpepm0xbba9pwe9Q8fs0-yqaqpepae9pg0FirpepeKkFr0xfr-x
% fr-xb9adbaqaaeGaciGaaiaabeqaamaabaabaaGcbaGaamyBamaaBa
% aaleaacaWGXbaabeaakiabg2da9iGacsgacaGGPbGaaiyyaiaacEga
% caGGOaWaaGbaaeaacaWGTbWaaSbaaSqaaiaadgeaaeqaaOGaaiilai
% aad2gadaWgaaWcbaGaamOqaaqabaaabaGaaeODaiaabggacaqGSbGa
% aeyzaiaab6gacaqGJbGaaeyzaaGccaGL44pacaGGSaWaaGbaaeaaca
% WGTbWaaSbaaSqaaiaaigdaaeqaaOGaaiilaiaad2gadaWgaaWcbaGa
% aGOmaaqabaGccaGGSaGaamyBamaaBaaaleaacaaIZaaabeaaaeaaca
% qGZbGaaeyzaiaabggaaOGaayjo+dGaai4oamaayaaabaGaamyBamaa
% BaaaleaacaWGbbaabeaakiaacYcacaWGTbWaaSbaaSqaaiaadkeaae
% qaaaqaaiaabEgacaqGObGaae4BaiaabohacaqG0baakiaawIJ-aiaa
% cMcaaaa!634E!
\begin{align}\eqn{PQmass}
m_q  = \operatorname{diag} (\underbrace {m_A ,m_B }_{\text{valence}},\underbrace {m_1 ,m_2 ,m_3 }_{\text{sea}};\underbrace {m_A ,m_B }_{\text{ghost}}).
\end{align}
% MathType!End!2!1!
\CPT~for PQ QCD is constructed in terms  of this matrix, or
in terms of $\chi$ which is still defined through \Eq{chidef}, and the
result is a Lagrangian identical to the one for ordinary \CPT, but
with an extended flavor structure and with super-traces replacing the traces.
Because of the great formal similarity between QCD and
PQ QCD, the extension of PQ \CPT~to PQ W\CPT~is a simple
generalization of the discussion in the previous section.
In particular, the
LO and NLO Lagrangians for PQ W\CPT~have forms
just like in \Eqsii{L2}{L4}, with traces replaced by
super-traces. Further, as in the continuum PQ \CPT,
the low-energy constants $L_i$'s and $W_i$'s are exactly the same in
the unquenched and PQ W\CPT. It follows that one can use the Wilson chiral
expressions for PQ theories to extract the GL coefficients. 

For completeness, we provide the expressions for the mass and decay 
constant for
PQ W\CPT~ in Appendix~\ref{PQresults}.
Again, as in unquenched theories, the LO and NLO loop
results in PQ W\CPT~are trivially related to the corresponding results in PQ
\CPT~ which have been calculated in \cite{SharpeShoresh2000}. 

%%%%%%%%%%%%%%%%%%%%%%%%%%%%%%%%%%%%%%
\end{section}
\begin{section}{Summary}
\label{summary}
%%%%%%%%%%%%%%%%%%%%%%%%%%%%%%%%%%%%%%
We constructed a low-energy  EFT, W\CPT, of the
Wilson lattice action close to the continuum. The theory extends
\CPT, and the perturbative
framework is described in terms of two small
parameters --- the quark mass $m_q$ and the lattice spacing $a$. The
Gasser-Leutwyler chiral Lagrangian (through $\mcal{O}(p^4)$ in \CPT) was
modified to incorporate all linear dependence on $a$. We applied this theory to
calculate light meson masses and decay constants. The resulting
expressions capture all the linear dependence on $a$ as well as
non-trivial logarithms that entangle $a$ and $m_q$.
A useful application of this theory is the determination of the
Gasser-Leutwyler coefficients of ordinary \CPT~from lattice
simulations at small but finite $a$. 

%%%%%%%%%%%%%%%%%%%%%%%%%%%%%%%%%%%%%
\end{section}
\begin{acknowledgments}
%%%%%%%%%%%%%%%%%%%%%%%%%%%%%%%%%%%%
We thank S. Sharpe for many useful 
discussions during various stages of this work. G.R. would like to thank
R. Woloshyn for many helpful conversations. 
We acknowledge support in part by the Natural Sciences and Engineering Research
Council of Canada, and U.S. DOE grants No DE-AC03-76SF00098, 
DE-FG03-96ER40956/A006 and DE-FG02-91ER40676.  

%%%%%%%%%%%%%%%%%%%%%%%%%%%%%%%%%%%%%%
\end{acknowledgments}
\appendix
\begin{section}{\CPT~results}
\label{ordinaryCPT}
%%%%%%%%%%%%%%%%%%%%%%%%%%%%%%%%%%%%%%%
We present the expressions for the mass of a flavor-charged light meson in 
\CPT~for comparison with W\CPT~ results. As explained in the text, one can see 
that the LO and NLO loop expressions
in W\CPT~can be obtained from the corresponding \CPT~results with the
substitution $\chi\rightarrow\chi+\rho$ ($\chi\to\mu$).  
Using the same notation as in \Eqs{WmassLO}{WR}, 
the masses through NLO with three quark flavors are~\cite{SharpeShoresh2000}:
\beq
M_{AB}^2=(M_{AB}^2)_\mathrm{LO}+(M_{AB}^2)_\mathrm{NLO,\
  loop}+(M_{AB}^2)_\mathrm{NLO,\ tree}\ ,
\eeq
with 
\begin{align}
\left( {M_{AB}^2 } \right)_{\text{LO}}  &  = \chi _{AB} \eqn{massLO}\ , \\ 
  \left( {M_{AB}^2 } \right)_{\text{NLO, loop}}  &  = \frac{1}
{{48f^2 \pi ^2 }}\chi _{AB} \sum\limits_{x = \pi ,\eta } {R_x^{AB} \chi _x \log
  \chi _x } \eqn{massNLOloop}\ , \\ 
  \left( {M_{AB}^2 } \right)_{\text{NLO, tree}}  &  = \frac{{24}}
{{f^2 }}\left( {2L_6  - L_4 } \right)\chi _{AB} \bar \chi  + \frac{8}
{{f^2 }}\left( {2L_8  - L_5 } \right)\chi _{AB}^2 \eqn{massNLOtree}\ ,
\end{align}
% MathType!End!2!1!
where
% MathType!Translator!2!1!Noam LaTeX.tdl!Noam -- AMS-LaTeX!
% MathType!MTEF!2!1!+-
% feaafaart1ev1aaatCvAUfeBSjuyZL2yd9gzLbvyNv2CaerbuLwBLn
% hiov2DGi1BTfMBaeXatLxBI9gBaerbd9wDYLwzYbItLDharqqtubsr
% 4rNCHbGeaGqiVu0Je9sqqrpepC0xbbL8F4rqqrFfpeea0xe9Lq-Jc9
% vqaqpepm0xbba9pwe9Q8fs0-yqaqpepae9pg0FirpepeKkFr0xfr-x
% fr-xb9adbaqaaeGaciGaaiaabeqaamaabaabaaGceaqabeaacqaHhp
% WydaWgaaWcbaGaeqiWdahabeaakiabgUcaRiabeE8aJnaaBaaaleaa
% cqaH3oaAaeqaaOGaeyypa0JaaGOmaiqbeE8aJzaaraaabaGaeq4Xdm
% 2aaSbaaSqaaiabec8aWbqabaGccqaHhpWydaWgaaWcbaGaeq4TdGga
% beaakiabg2da9iabeE8aJnaaBaaaleaacaaIXaaabeaakiabeE8aJn
% aaBaaaleaacaaIYaaabeaakiabgUcaRiabeE8aJnaaBaaaleaacaaI
% XaaabeaakiabeE8aJnaaBaaaleaacaaIZaaabeaakiabgUcaRiabeE
% 8aJnaaBaaaleaacaaIYaaabeaakiabeE8aJnaaBaaaleaacaaIZaaa
% beaaaOqaaiaadkfadaqhaaWcbaGaeqiWdahabaGaamyqaiaadkeaaa
% GccqGH9aqpdaWcaaqaaiabeE8aJnaaBaaaleaacaWGdbaabeaakiab
% gkHiTiabeE8aJnaaBaaaleaacqaHapaCaeqaaaGcbaGaeq4Xdm2aaS
% baaSqaaiabeE7aObqabaGccqGHsislcqaHhpWydaWgaaWcbaGaeqiW
% dahabeaaaaGccaGGSaGaaGzbVlaadkfadaqhaaWcbaGaeq4TdGgaba
% GaamyqaiaadkeaaaGccqGH9aqpdaWcaaqaaiabeE8aJnaaBaaaleaa
% caWGdbaabeaakiabgkHiTiabeE8aJnaaBaaaleaacqaH3oaAaeqaaa
% GcbaGaeq4Xdm2aaSbaaSqaaiabec8aWbqabaGccqGHsislcqaHhpWy
% daWgaaWcbaGaeq4TdGgabeaaaaGccaaMc8UaaiOlaaaaaa!89D9!
\begin{align}
& \chi _\pi   + \chi _\eta   = 2\bar \chi\ ,  \\ 
 & \chi _\pi  \chi _\eta   = \chi _1 \chi _2  + \chi _1 \chi _3  + \chi _2 \chi
 _3\ ,  \\ 
 & R_\pi ^{AB}  = \frac{{\chi _C  - \chi _\pi  }}
{{\chi _\eta   - \chi _\pi  }},\quad R_\eta ^{AB}  = \frac{{\chi _C  - \chi _\eta  }}
{{\chi _\pi   - \chi _\eta  }}\,.
\end{align}
% MathType!End!2!1!
(Here, again, $C$ is the flavor that is different from both $A$ and $B$.)

%%%%%%%%%%%%%%%%%%%%%%%%%%%%%%%%%%%%%%%%%%%%%%
\end{section}
\begin{section}{PQ W\CPT~results}
\label{PQresults}
%%%%%%%%%%%%%%%%%%%%%%%%%%%%%%%%%%%%%%%%%%%%%
The forms of the Lagrangians in unquenched and PQ theory are the same,
with an implicit difference in the structure of the matrices in 
flavor space and the replacement
of traces with super-traces. Thus all
tree contributions in both theories have the same dependence on $\chi$ and
$\rho$.   
In particular, the LO and NLO tree results are still given by 
\Eqsii{WmassLO}{WmassNLOtree} and 
\Eqsii{WfLO}{WfNLOtree} for the mass and the decay constant
respectively, with appropriate $\chi$ and $\rho$ matrices for PQ simulations
(the structure of $\rho$ in PQ W\CPT~ is determined by $r$ in the PQ
version of the Wilson action. The latter must have a structure similar
to $m_q$, \eq{PQmass}, that is needed to guarantee the exact
cancellation between valence and ghost loops.). We give here only the NLO
loop results, which are different from the unquenched expressions.  

\beq
(M_{AB}^2)_\mathrm{NLO,\ loop}\Align
\frac{1}{{16f^2 \pi ^2 N}}\ \mu_{AB} 
\sum\limits_{x = A,B,\pi ,\eta }
 {R_x \mu_x \log (\mu_x) }\ ,\\
(f_{AB})_\mathrm{NLO,\ loop}\Align - \frac{1}
{{64\pi ^2 f}}\sum\limits_{\begin{subarray}{l} 
  i = 1,2,3 \\ 
  j = A,B 
\end{subarray}}  {\mu _{ij} \log \mu _{ij} }+\frac{1}{192\pi^2f}
\left\{-D_A-D_B \phantom{+ \frac{\log(\mu_A/\mu_B)}{\mu_A-\mu_B}} 
\right.\nonumber\\
&{}&+ \frac{\log(\mu_A/\mu_B)}{\mu_A-\mu_B}
\left[\mu_A D_A + \mu_B D_B + (\mu_A-\mu_B)^2\right]\nonumber\\
&{}&\left. + \sum_{x=\pi,\ \eta} R_x\mu_x (\mu_A-\mu_B)
\left[\frac{\log(\mu_A/\mu_x)}{\mu_A-\mu_x} -
\frac{\log(\mu_B/\mu_x)}{\mu_B-\mu_x}\right] 
 \right\}\ ,
\eeq
where

\beq
R_x  = \frac{{\prod\limits_{ i = 1,2,3} {( \mu_i - 
\mu_x )}
 }}
{{\prod\limits_{ \begin{matrix}
{y = A,B,\pi ,\eta }\\
{y \ne x}\\
\end{matrix}} {(\mu_y  - \mu_x )} }}\ ,\ \ \ 
D_x = \frac{\prod\limits_{i=1,2,3}
(\mu_i-\mu_x)}{(\mu_\pi-\mu_x)(\mu_\eta-\mu_x)}
\, \ .
\eeq

%%%%%%%%%%%%%%%%%%%%%%%%%%%%%%%%%%%%%%%%%%%%%
\end{section}
%%%%%%%%%%%%%%%%%%%%%%%%%%%%%%%%%%%%%%%%%%%%%%%%%%%%%%%%%%%%%
%\include{Bibliography}
%\bibliographystyle{unsrt} 

%\bibliographystyle{apsrev}


\begin{thebibliography}{10}

\bibitem{SharpeSingleton1998}
S.~R. Sharpe and R. Singleton Jr., \Journal{\PRD}{58}{074501}{1998}.

\bibitem{Luscher}
M. Luscher, S. Sint, R. Sommer and P. Weisz, \Journal{\NPB}{478}{365}{1996}.

\bibitem{Symanzik83}
K. Symanzik, \Journal{\NPB}{226}{187}{1983}.

\bibitem{LuscherWeisz96}
M. Luscher and P. Weisz, \Journal{\NPB}{479}{429}{1996}, M.~Luscher, et.
  al.,~\Journal{\NPB}{491}{323}{1997}.

\bibitem{BGLS2001}
T. Bhattacharya, et. al., {\it Nucl. Phys. Proc. Suppl.}~{\bf
  106},~{789}~{(2001)}.

\bibitem{morel87}
A. Morel, {\it J. Phys}\ (France)\ {\bf 48}, 1111 (1987).

\bibitem{Bernard}
C.~W. Bernard and M.~F.~L. Golterman, \Journal{\PRD}{49}{486}{1994}.

\bibitem{SharpeShoresh2001}
S.~R. Sharpe and N. Shoresh, \Journal{\PRD}{64}{114510}{2001}.

\bibitem{SharpeShoresh2000}
S.~R. Sharpe and N. Shoresh, \Journal{\PRD}{62}{094503}{2000}.

\bibitem{CohenKaplanNelson99}
A.~G. Cohen, D.~B. Kaplan and A.~E. Nelson, {\em JHEP}, 11:027, 1999.

\end{thebibliography}
\end{document}